\documentclass[showkeys,12pt,twocolumn]{article}

\usepackage{arxiv}
\usepackage[english]{babel}
\usepackage[utf8]{inputenc}
\usepackage[backend=bibtex,style=phys,biblabel=brackets,articletitle=false]{biblatex}
\usepackage{xcolor} 
\usepackage[colorinlistoftodos, color=green!40, prependcaption]{todonotes}
\usepackage{mathrsfs}
\usepackage{amsmath}
\usepackage{amsfonts} 
\usepackage{amssymb}
\usepackage{amsthm}
\usepackage{mathtools}  
\usepackage{slashed}
\usepackage{enumitem}
\usepackage[normalem]{ulem}
\usepackage{lipsum}
\usepackage[colorlinks=true,linkcolor=blue,urlcolor=blue,citecolor=blue]{hyperref}
\usepackage[capitalise]{cleveref}

\makeatletter
\newcommand{\mybf}[1]{%
    \begingroup
    \def\@tempa{#1}%
    \ifx\@tempa\@empty
        \textbf{#1}%
    \else
        \edef\@tempa{\noexpand\in@{#1}{abcdefghijklmnopqrstuvwxyzABCDEFGHIJKLMNOPQRSTUVWXYZ}}%
        \@tempa
        \ifin@
            \mathbf{#1}%
        \else
            \pmb{#1}%
        \fi
    \fi
    \endgroup
}
\makeatother

\makeatletter
\newif\if@insideout
\@insideoutfalse

\DeclareRobustCommand{\mysout}[1]{%
  \begingroup
  \@insideouttrue
  \ifmmode
    \expandafter\cancel
  \else
    \expandafter\sout
  \fi{#1}%
  \@insideoutfalse
  \endgroup
}

\newcommand{\citepatch}[1]{\begingroup\color{red}\sout{\oldcite{#1}}\endgroup}

\let\oldcite\cite
\renewcommand{\cite}[1]{%
  \if@insideout
    \citepatch{#1}%
  \else
    \oldcite{#1}%
  \fi
}
\makeatother

\begin{document}

\title{Thermal wakefield structure in plasma acceleration processes: insights from fluid models and PIC simulations}

\author{
     Daniele Simeoni$^{1,*}$
\And Andrea Renato Rossi$^{2}$
\And Gianmarco Parise$^{3}$
\And Fabio Guglietta$^{1}$
\And Mauro Sbragaglia$^{1}$
\and \\
$^{1}$Department of Physics \& INFN, Tor Vergata University of Rome, Via della Ricerca Scientifica 1, 00133, Rome, Italy
\and
$^{2}$INFN, Section of Milan, via Celoria 16, 20133, Milan, Italy
\and
$^{3}$INFN, Laboratori Nazionali di Frascati, Via Enrico Fermi 54, 00044, Frascati, Italy
\and
$^{*}$\texttt{daniele.simeoni@roma2.infn.it}
}

\date{\today} 

\twocolumn[
\maketitle

\begin{abstract}
We focus on the process of plasma acceleration in the presence of non-negligible thermal effects, wherein a driver of relativistic electrons perturbs a warm neutral plasma and generates a wakefield structure. We study the acceleration process via numerical simulations based on fluid models with different thermal closure assumptions, and also provide systematic comparisons against ground-truth data coming from particle-in-cell (PIC) simulations. The focus of the analysis is on the first electron depletion bubble after the driver, where we provide a detailed characterization of its size and the electromagnetic fields developed inside. Our results are instrumental in determining the correct thermal closure assumption to be used in fluid models for the numerical simulations of plasma acceleration processes, as well as elucidating the corresponding limits of applicability.
\end{abstract}
]

\section{Introduction}
\label{sec:intro}
Plasma wakefield acceleration (PWFA) is a novel technique~\cite{
joshi-2007,tajima-2020,ferrario-2021} that relies on the interaction between a relativistic bunch of electrons (driver) and a background neutral plasma. As the driver moves through the plasma, an oscillating structure is created due to the Coulomb repulsion force, and this particular configuration -- \textit{the wakefield} -- favors strong electromagnetic fields in selected regions of the wake, thus producing accelerating forces that are orders of magnitude larger than the ones obtained by conventional acceleration techniques~\cite{joshi-2003,schroeder-2020,kurz-2021,shiltsev-2021}.
The physical scenario under consideration is a complex and intrinsically multiscale problem and, in this context, numerical simulations have been established as an essential tool to complement experimental investigation and to deepen understanding of the underlying physics~\cite{birdsall-2018}. 
In the early stages of plasma acceleration, the relevant time scales are so short that particle collisions can be neglected; the plasma dynamics is therefore successfully captured by the relativistic Vlasov-Maxwell system~\cite{nicholson-1983,landau-1987,cercignani-2002} that solves for the particles' distribution functions. 

\begin{figure}[h!]
    \includegraphics[width=\columnwidth]{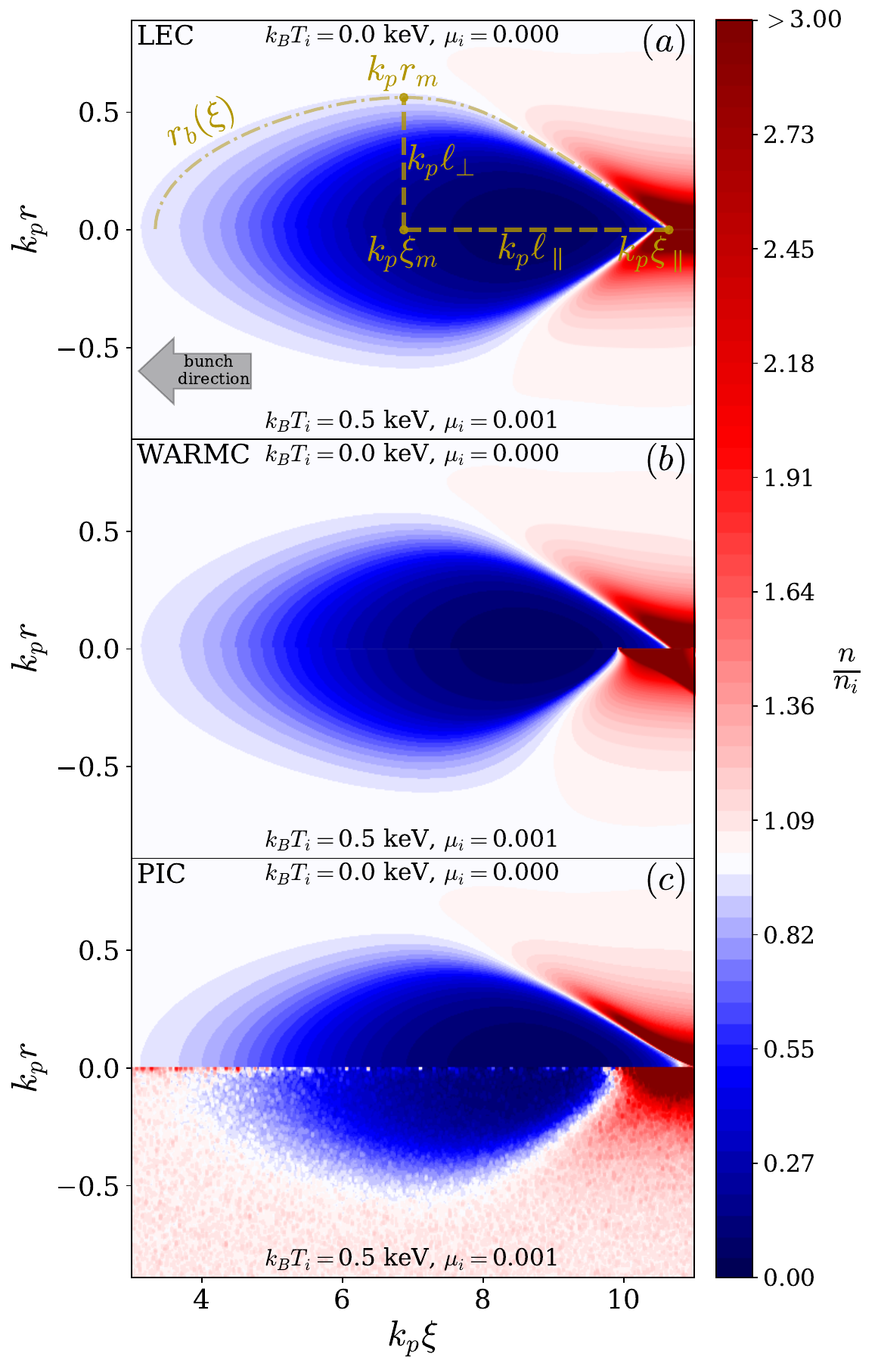}
    \caption{\footnotesize{Snapshots of normalized electron plasma density $n/n_i$ for LEC (panel (a)), WARMC (panel (b)) and PIC (panel (c)). Each panel shows a comparison between the cold case ($k_B T_i = 0~\rm{keV}$, upper half) and a warm case ($k_B T_i = 0.5~\rm{keV}$, lower half). Corresponding values of $\mu_i = k_B T_i/m_e c^2$ are also reported. In panel (a), a sketch is provided to illustrate the longitudinal ($\ell_{\parallel}$) and the transverse ($\ell_{\bot}$) size of the bubble (see text for details). Simulations are performed with $\tilde{Q}= 1.0$. Spatial coordinates are made dimensionless w.r.t. $k_p^{-1}$.}}
    \label{fig:1}
\end{figure}

Particle-in-cell (PIC) numerical methods~\cite{birdsall-2018,hockney-2021} solve the relativistic Vlasov-Maxwell equations by advancing the trajectories of relativistic charged particles under the effect of electromagnetic fields. The resulting macroscopic fields (density, momentum, etc.) are consequently reconstructed from the particle ensemble, which inevitably leads to statistical noise. Reducing the noise requires a large number of particles, and thus significantly increases the computational cost~\cite{kesting-2015,tavassoli-2021}.
Other numerical approaches rely on coarse-grained descriptions of the relativistic Vlasov-Maxwell system, yielding a fluid description of the plasma~\cite{landau-1987}. From the technical point of view, this amounts to taking the hierarchy of equations for the moments of the kinetic distribution function, obtained from the Vlasov-Maxwell system, and then applying a suitable closure scheme to truncate the system~\cite{krall-1973}. Inevitably, the resulting fluid equations are not fully equivalent to the underlying kinetic description; hence, while fluid models avoid statistical noise by construction, they may fail to capture certain kinetic effects~\cite{vlasov-1961,landau-1965,shalchi-2021}.
This calls for studies to understand the limitations of fluid models and identify the conditions under which such models provide a faithful description of plasma acceleration processes. With this aim, recent studies provided detailed comparisons between fluid models and PIC simulations~\cite{massimo-2016-a,benedetti-2010}; however such comparisons have been mainly limited to cold closures, i.e., situations where the plasma temperature can be neglected. Although thermal effects were already invoked many years ago as a possible means of regularization for the wakefield singularity close to wavebreaking~\cite{katsouleas-1988}, they are frequently neglected in actual simulations of PWFA, on the tenet that thermal energy of the plasma is ordders of magnitude smaller than the particle rest energy~\cite{simeoni-2024}.
Recent research studies, however, highlighted a non trivial role of thermal effects in plasma acceleration contexts. For example, thermal effects and cumulative heating of the plasma in high repetition rate contexts could modify the wakefield structure~\cite{gholizadeh-2011,gilljohann-2019,darcy-2022}; thermal effects could impact the ion channel formation~\cite{zgadzaj-2020,khudiakov-2022} and also bear some improvement on the beam quality in positron based accelerators~\cite{silva-2021,diederichs-2023,diederichs-2023-b,cao-2024}. Correspondingly, a possible use of fluid models with non-negligible thermal effects poses the problem of which closure assumption to adopt in the kinetic equations~\cite{toepfer-1971, newcomb-1982, amendt-1985, newcomb-1986-b, siambis-1987, pennisi-1991,muscato-1993}. Plasma dynamics in the early stages of PWFA is essentially collisionless, hence the application of a {\it local equilibrium closure} (LEC) based on the assumption that the kinetic distribution function is close to a local equilibrium does not seem appropriate. From the computational side, however, one has to notice that the number of fluid equations with LEC is limited, since one has to consider only mass and isotropic momentum equations~\cite{toepfer-1971,cercignani-2002,rezzolla-2013}; hence, if the associated modeling error was controllable, LEC descriptions could still be in principle a valid tool of analysis for problems where the parameters space is large, and fast tools are needed for preliminary investigations. Other closures are indeed possible: this is the case of the 
\textit{warm closure} (WARMC), based on the assumption of small thermal spread without any constraints for the kinetic distribution function to be close to a local equilibrium~\cite{amendt-1985, newcomb-1982, siambis-1987,schroeder-2005, schroeder-2009, schroeder-2010}. Fluid models based on WARMC, however, result in an increased  number of fluid equations, larger than the ones obtained with LEC; still, they are supposed to be closer to the physics of kinetic equations. In particular, with the WARMC, the stress tensor is found to be anisotropic~\cite{shadwick-2004,shadwick-2005} whereas in the LEC model is not. 
In~\cite{simeoni-2024-b}, by systematic comparisons between WARMC/LEC and PIC spatially resolved simulations, we have indeed shown that such a feature of anisotropy is well reproduced with WARMC. While the measure of the stress tensor anisotropy already provides a meaningful indicator of the quality of the thermal closure, it remains an indirect diagnostic. In this work, we significantly extend the analysis to observables that can be more directly related to the efficiency and quality of the acceleration process. In particular, we refer to the geometry of the first electron depletion bubble that is formed after the driver, characterized through its longitudinal and transverse size (see \cref{fig:1}), and to the accelerating and focusing wakefields that develop inside it. These quantities allow us to test the robustness of the WARMC and LEC thermal closures, which are both theoretically justified only for small thermal spreads, as functions of the initial background plasma temperature, and to give a quantitative assessment of the quality of their description as well as their limitations w.r.t. fully kinetic PIC descriptions. To the best of our knowledge, this study is the first systematic assessment of thermal effects on the wakefield structure of PWFA obtained through spatially resolved fluid simulations directly compared against PIC data.
The paper is organized as follows.~\cref{sec:Vlasov} recalls the basic features of the relativistic Vlasov-Maxwell equations.~\cref{sec:PIC,sec:fluids} describes the methods employed to model and numerically solve the relativistic kinetic equations.~\cref{sec:results} discusses our results, and conclusions are drawn in~\cref{sec:conclusions}.
\begin{figure}
\includegraphics[width=\columnwidth]{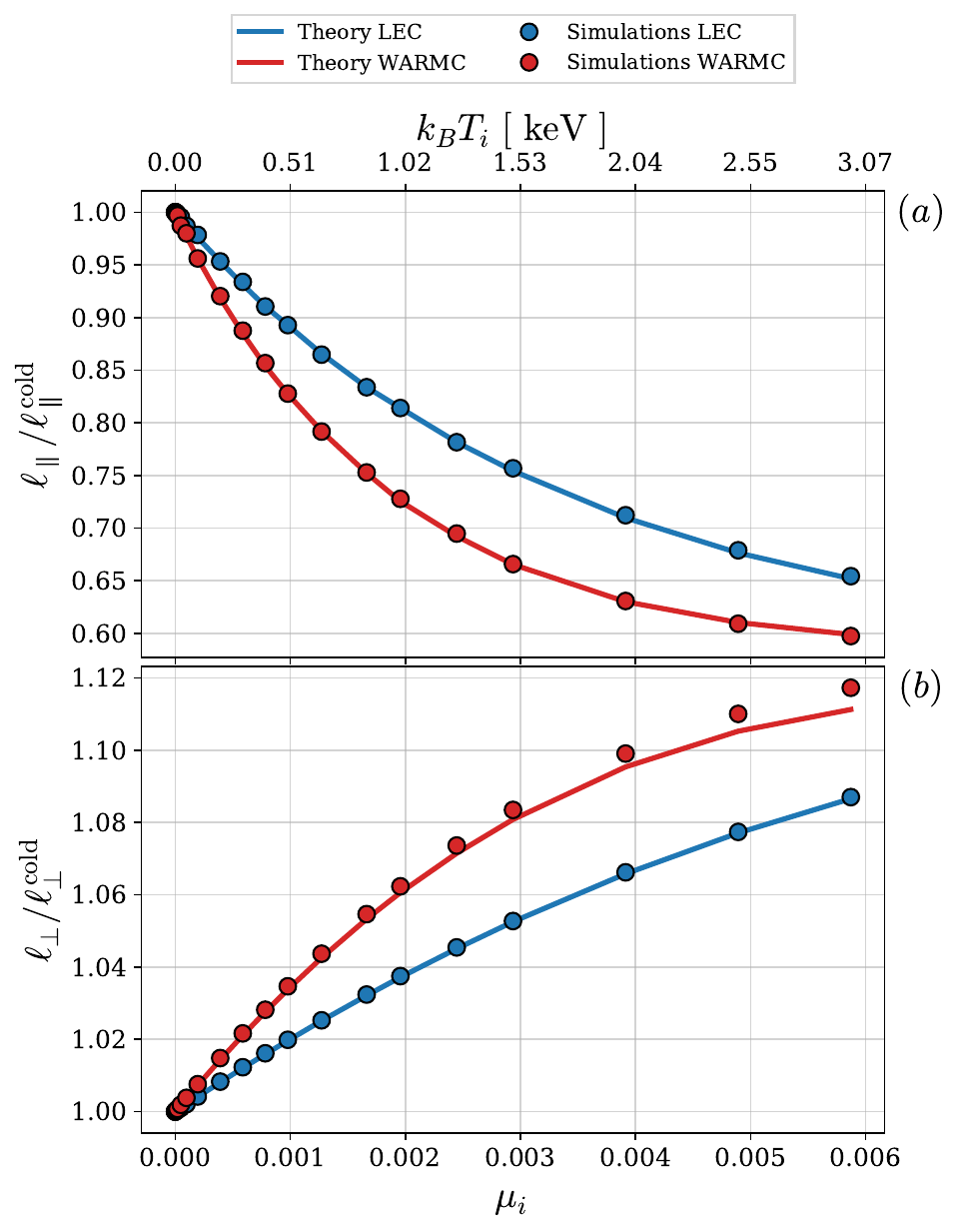}
    \caption{\footnotesize{We analyze $\ell_{\parallel}$ and $\ell_{\bot}$ at changing $k_B T_i$ (or equivalently $\mu_i = k_B T_i/m_e c^2$) for LEC (blue points/lines) and WARMC (red points/lines) at fixed $\tilde{Q}=10^{-4}$. Results of numerical simulations (points) are compared with corresponding theoretical predictions in the linear regime (lines, see text for details). Both $\ell_{\parallel}$ and $\ell_{\bot}$ are normalized w.r.t. the corresponding value in the cold limit ($\ell^{\rm{cold}}_{\parallel}$, $\ell^{\rm{cold}}_{\bot}$).}}
    \label{fig:2}
\end{figure}
%
\section{Background: Relativistic Vlasov-Maxwell Equations}
\label{sec:Vlasov}
%
The following section briefly recalls the fundamental equations underlying the PWFA phenomenon. A more comprehensive treatment can be found in~\cite{cercignani-2002}. Throughout this work, we adopt when needed a manifestly covariant formulation and Einstein’s summation convention, with Greek indices denoting space–time coordinates.

The fundamental building block for the theoretical description of a warm gas of electrons with mass $m_e$ and charge $-e$ is the Vlasov equation~\cite{vlasov-1961}:
\begin{align}\label{eq:vlasov-eq}
p^{\alpha} \partial_\alpha f - \frac{e}{c} F^{\alpha\beta} p_{\beta} \frac{\partial f}{\partial p^\alpha} = 0    \; ,
\end{align}
describing the conservation of the density function $f=f(x^{\alpha}, p^{\alpha})$ in the 7-dimensional phase space of electron's space-time coordinates $x^{\alpha}=(ct, \mybf{x})$ ($c$ being the speed of light) and relativistic kinetic momenta $p^{\alpha}=(p^0, \mybf{p})$. Here, $\partial_\alpha = \frac{\partial}{\partial x^\alpha}$ is the space-time derivative. From the density function, one can build its low order moments:
\begin{equation}
    \label{eq:moments}
    \small
    {
    \begin{aligned}
    \text{\textit{invariant density:}}      \;
    &h                       = c \int f \frac{d \mybf{p}}{p^0} \;, \\
    \text{\textit{particle flow:}}          \;
    &N^{\alpha}              = c \int f p^{\alpha} \frac{d \mybf{p}}{p^0} \;,\\
    \text{\textit{energy-momentum tensor:}} \;
    &T^{\alpha \beta}        = c \int f p^{\alpha} p^{\beta} \frac{d \mybf{p}}{p^0} \;,\\
    \text{\textit{energy-momentum flux:}} \;
    &M^{\alpha \beta \gamma} = c \int f p^{\alpha} p^{\beta} p^{\gamma} \frac{d \mybf{p}}{p^0} \ ,
    \end{aligned}
    }
\end{equation}
whose governing equations stem from~\cref{eq:vlasov-eq} as an expression of mass, momentum, and energy conservation:
\begin{align}
\label{eq:cons_eqs1}
0 &= \partial_\alpha N^{\alpha}                                               \; , \\
\label{eq:cons_eqs2}
0 &= \partial_\alpha T^{\alpha\beta} + \frac{e}{c} F^{\beta\alpha} N_{\alpha} \; , \\
\label{eq:cons_eqs3}
0 &= \partial_\alpha M^{\alpha\beta\gamma}+\frac{e}{c}(F^{\beta\alpha}T_{\alpha}^{~\gamma}+F^{\gamma\alpha}T_{\alpha}^{~\beta})     \; . 
\end{align}
These are the core equations that are needed for a relativistic hydrodynamic description of a PWFA system, provided that they are coupled with Maxwell's equations:
\begin{align}
    \label{eq:maxwell-inhomogeneus}
    0 &= \partial_\alpha F^{\alpha\beta} + \mu_0 c e 
    (N^{\beta}_{b} + N^{\beta} - N^{\beta}_{i}) \;,  \\
    \label{eq:maxwell-homogeneus}
    0 &= \partial_\alpha F_{\beta\gamma} + \partial_\beta F_{\gamma\alpha} + \partial_\gamma F_{\alpha\beta} \;,  
\end{align}
which describe the evolution of the electromagnetic field tensor $F^{\alpha\beta}$ (whose components are the electric and magnetic fields, $\mybf{E}$ and $\mybf{B}$, respectively and the subscript $b$ represents the bunch quantities)~\cite{jackson-1998} and that read in non-covariant form as:
\begin{align}
    \label{eq:maxwell-1}
    & \nabla \cdot  \mybf{E} = -\frac{e}{\epsilon_0} (n_b + n - n_i)                 \;,  \\
    \label{eq:maxwell-2}
    & \nabla \cdot  \mybf{B} = 0                                                     \;,  \\
    \label{eq:maxwell-3}
    & \nabla \wedge \mybf{E} = -\partial_t \mybf{B}                                  \;,  \\
    \label{eq:maxwell-4}
    & \nabla \wedge \mybf{B} = - \mu_0 e (n \mybf{u} + n_b \mybf{u}_b)  + \epsilon_0 \mu_0 \partial_t \mybf{E} \;,
\end{align}
where $\epsilon_0$ and $\mu_0$ are respectively the vacuum's permittivity and permeability. Here:
\begin{align}
    N^{\beta}_{b} = n_b
    \begin{pmatrix}
        c           \\
        \mybf{u}_b  
    \end{pmatrix} , \; \;
    N^{\beta}    = n
    \begin{pmatrix}
        c           \\
        \mybf{u}  
    \end{pmatrix} , \; \;
    N^{\beta}_{i} = n_i
    \begin{pmatrix}
        c           \\
        \mybf{0}  
    \end{pmatrix} \; ,
\end{align}
are respectively the contributions coming from an ultra-relativistic ($\mybf{u}_b=-c\hat{\mybf{z}}$) driving electron bunch with Gaussian number density $n_b$, and from the background plasma electrons, with number density $n$ and Eulerian velocity field $\mybf{u}$. $n_i$ is the initial uniform rest number density of plasma electrons, equal (due to charge neutrality) to the constant plasma ion's number density: ions are considered immobile in our treatment.
%
\section{Methods: PIC Simulations}
\label{sec:PIC}
Direct numerical simulation of~\cref{eq:vlasov-eq} is a formidable task, as it requires the discretization of a 7-dimensional phase space. Such an endeavor remains computationally demanding even on modern high-performance computers. Consequently, the standard approach to the simulation of PWFA systems has been represented by PIC codes~\cite{birdsall-2018}, which model the dynamics at the microscopic level of individual particles, by integrating the following equations of motion: 
\begin{figure}
\includegraphics[width=\columnwidth]{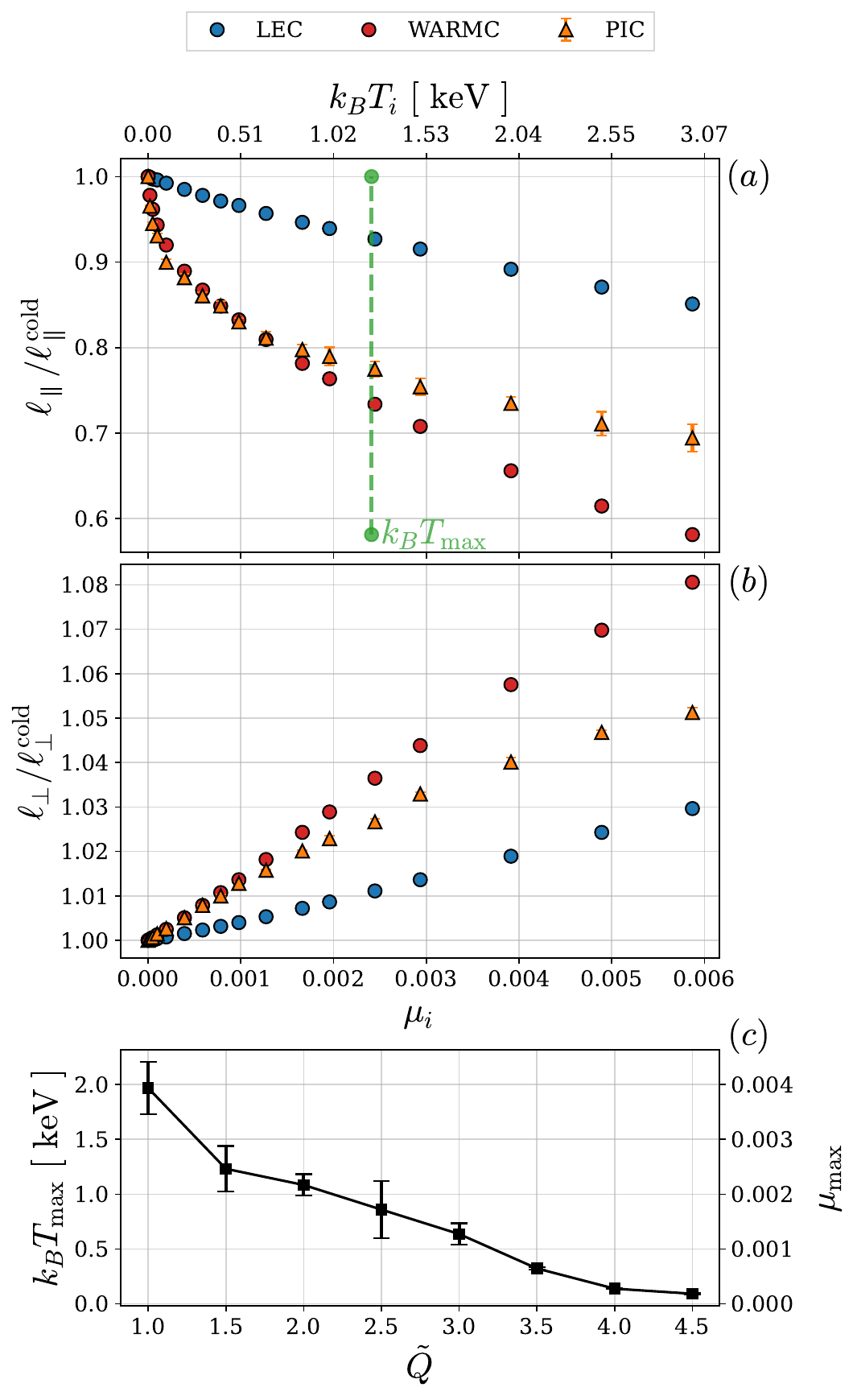}
    \caption{\footnotesize{Panels (a) and (b): we analyze $\ell_{\parallel}$ and $\ell_{\bot}$ at changing $k_B T_i$ (or equivalently $\mu_i = k_B T_i/m_e c^2$) for LEC (blue points), WARMC (red points) and PIC (orange triangles) data at fixed $\tilde{Q}=1.5$. Both $\ell_{\parallel}$ and $\ell_{\bot}$ are normalized w.r.t. the corresponding value in the cold limit ($\ell^{\rm{cold}}_{\parallel}$, $\ell^{\rm{cold}}_{\bot}$). Error-bars in PIC data are obtained by averaging 9 independent simulations. Panel (c): we analyze $k_B T_{\rm{max}}$ (or equivalently $\mu_{\rm{max}}$) at changing $\tilde{Q}$. $T_{\rm{max}}$ is the temperature at which the relative discrepancy in $\ell_\parallel$ between WARMC and PIC reaches a threshold value (see text for details).}
    }
    \label{fig:3}
\end{figure}
\begin{align}
    \begin{cases}
    \frac{d\mybf{x}}{dt} &= \mybf{v}                                                \\
    \frac{d\mybf{p}}{dt} &= -e \left( \mybf{E}+\mybf{v} \wedge \mybf{B} \right)     \\
    \mybf{p}             &= m_e \mybf{v} / \sqrt{1-(\mybf{v}/c)^2}                
    \end{cases} \; ,
\end{align}
via the adoption of leap-frog integrators such as the Boris~\cite{boris-1970}, Vay~\cite{vay-2008}, or Cary-Higuera~\cite{higuera-2017} particle pushers. A comprehensive review on these schemes can be found in~\cite{ripperda-2018}. 
Nevertheless, simulating all the electrons in the plasma individually is unfeasible; therefore, PIC methods group electrons into macro-particles, each representing a large ensemble of real electrons sharing the same position and velocity. This under-sampling introduces statistical noise, which constitutes the main source of uncertainties in these schemes. 
In addition to solving Maxwell's curl~\cref{eq:maxwell-3,eq:maxwell-4} on a fixed Eulerian grid, PIC algorithms must ensure coupling between Lagrangian particles and grid-based electromagnetic fields, and they often do so by using B-spline shape functions for the interpolation, or more sophisticated charge-conserving methods~\cite{esirkepov-2001,umeda-2003,steiniger-2023}. In this framework, an initial background temperature is set by simply imposing a variance in the particles' initial momenta.\\
In this work, we employ the FBPIC code~\cite{lehe-2016}, a quasi-3D PIC framework that uses a Hankel transform to decompose fields and currents in the transverse direction (with the maximum azimuthal mode order specified by the user), thereby extending the method beyond the constraint of pure cylindrical symmetry. FBPIC also employs a spectral Maxwell solver that eliminates numerical dispersion for a relativistic, near-speed-of-light dynamic.
\section{Methods: Fluid Models}
\label{sec:fluids}
At variance with PIC schemes, which provide a microscopic description of the plasma to solve~\cref{eq:vlasov-eq}, fluid models adopt a macroscopic perspective, solving for the moments of the phase-space distribution function~\cref{eq:moments}.
Their main advantage lies in the absence of the statistical noise that arises from particle undersampling in PIC simulations: fluid equations inherently average over microscopic scales by construction. However, this same averaging inevitably entails the loss of some physics~\cite{nicholson-1983}.\\
The constitutive equations of fluid models are the conservation~\cref{eq:cons_eqs1,eq:cons_eqs2,eq:cons_eqs3}, 
which form an unclosed system and therefore require a closure relation. The most common choice in the PWFA community has traditionally been the \textit{cold closure}, which assumes a negligible plasma temperature. Under this approximation, the particle flow and energy-momentum tensor assume the following form:
\begin{align}
    N^{\alpha} = n_0 U^{\alpha} \;, \quad T^{\alpha\beta} = n_0 m_e U^{\alpha}U^{\beta} \;,
\end{align}
where $U^{\alpha}=\gamma (c,\mybf{u})$ is the fluid velocity, $n_0 = n / \gamma$ is the particle number density in the fluid rest frame (from now on, we will indicate all such quantities with the $_0$ subscript) and $\gamma=(1-(\mybf{u}/c)^2)^{-1/2}$ is the Lorentz factor associated with $\mybf{u}$.\\
When thermal effects become relevant, the cold closure is no longer valid, and alternative closure schemes must be adopted. In the following, we consider two popular \textit{thermal closure assumptions}, each expected to remain accurate for small thermal spreads. These closures are briefly reviewed in~\cref{sec:lec} and~\cref{sec:warmc}. The interested reader will find more details on their derivation in~\cite{muscato-1993}.\\
The fluid models resulting from the thermal closure assumptions are then numerically integrated via ad-hoc designed numerical algorithms based on the lattice Boltzmann equations~\cite{parise-2022,simeoni-2024}. A comprehensive discussion of the schemes used to numerically integrate these fluid models is presented in~\cite{simeoni-2024}.
\subsection{Local Equilibrium Closure (LEC)}
\label{sec:lec}
The first thermal closure considered in this work is the {\it Local Equilibrium Closure} (LEC), which assumes that the underlying fluid remains ideal, that is, it imposes that the distribution function is the equilibrium Maxwell-J{\"u}ttner distribution~\cite{juettner-1911}. Under this constraint, the particle flow $N^{\alpha}$ and energy-momentum tensor $T^{\alpha\beta}$ assume the following forms: 
\begin{align}\label{eq:ideal-tensors-fluid}
         N^{\alpha} = n_0 U^{\alpha} \; , \;
    T^{\alpha\beta} = (P_0+\varepsilon_0) \frac{U^\alpha U^\beta}{c^2} - P_0 \eta^{\alpha\beta} \; ,
\end{align}
with $P_0$, $\varepsilon_0$ the pressure and internal energy density respectively, and $\eta^{\alpha\beta}$ the Minkowsky metric tensor. Consequently, the conservation~\cref{eq:cons_eqs1,eq:cons_eqs2} can be rewritten as the relativistic analogue of Euler's equations, which we give here in non covariant form: 
\begin{equation}
    \label{eq:rel-euler}
    \begin{aligned}
    \frac{\partial n}{\partial t} + \nabla \cdot (n \mathbf{u}) &= 0 \; , \\
    \left[ \frac{\partial}{\partial t} + \mybf{u} \cdot \nabla \right] \left( \frac{h_0 \mybf{p}}{m_e c^2} \right) 
    &= - \frac{\nabla P_0}{n} - e (\mybf{E} + \mybf{u} \wedge \mybf{B}) \; ,
    \end{aligned}
\end{equation}
where $h_0$ is the relativistic enthalpy per particle.
Assuming the fluid is ideal allows one to adopt an ideal relativistic equation of state, namely, the Synge~\cite{synge-1957} equation of state in the small temperature limit, and this leads to a specific scaling relation for temperature $T$: 
\begin{align}
    \label{eq:lec-temp-scaling}
    T      &= T_i \left(\frac{n_0}{n_i}\right)^{2/3} \; , 
\end{align}
where $T_i$ denotes the initial background temperature in the plasma. The corresponding pressure is then given by $P_0 = n_0 k_B T$, with $k_B$ the Boltzmann constant, while enthalpy as $h_0 = m_e c^2 + (5/2) P_0 / n_0$. 
%
\subsection{Warm Plasma Closure (WARMC)} \label{sec:warmc}
A second thermal closure assumption is provided by the so called {\it Warm Closure} (WARMC)~\cite{schroeder-2010}. In a nutshell, the key idea is to re-express~\cref{eq:cons_eqs1,eq:cons_eqs2,eq:cons_eqs3} through the \textit{centered} moments of second ($\theta^{\alpha\beta}$) and third order ($Q^{\alpha\beta\gamma}$):
\label{eq:theta-integrals}
{
\begin{align}
    \theta^{\alpha\beta} &= c \int f 
    \left(p^{\alpha}-\frac{N^{\alpha}}{h}\right) 
    \left(p^{\beta}-\frac{N^{\beta}}{h}\right) 
    \frac{d \mybf{p}}{p^0} \;, \\
    Q^{\alpha\beta\gamma} &= c \int f 
    \left(p^{\alpha}-\frac{N^{\alpha}}{h}\right)
    \left(p^{\beta}-\frac{N^{\beta}}{h}\right)
    \left(p^{\gamma}-\frac{N^{\gamma}}{h}\right)
    \frac{d \mybf{p}}{p^0} \;, \notag
\end{align}
}
where one can easily verify that $\theta^{\alpha\beta} = T^{\alpha\beta} - N^{\alpha}N^{\beta}/h$. The closure is indeed performed by neglecting $Q^{\alpha\beta\gamma}$ in the conservation equations, on the excuse of small thermal spreads in the plasma. This leads to the following set of equations:
\begin{equation}\label{eq:warmc-fluid-eqs}
{
\begin{aligned}
      \partial_\alpha N^\alpha                                    &= 0 \;, \\
      N^\alpha \partial_\alpha \left( \frac{N^{\beta}}{h} \right) &= 
         - \partial_\alpha \theta^{\alpha\beta} 
         - \frac{e}{c}F^{\beta\alpha}N_{\alpha}                        \;, \\
      N^\alpha \partial_\alpha \left( \frac{\theta^{\beta\gamma}}{h} \right) &=  
          - \theta^{\gamma\alpha} \partial_\alpha \left( \frac{N^{\beta}}{h} \right)   
          - \theta^{\alpha \beta} \partial_\alpha \left( \frac{N^{\gamma}}{h} \right)  + \\
         &- \frac{e}{c}(F^{\beta\alpha}\theta_{\alpha}^{~\gamma}+F^{\gamma\alpha}\theta_{\alpha}^{~\beta})\;,
\end{aligned}
}
\end{equation}
which have to be coupled with the constraints coming from the mass-shell condition for kinetic momenta, $p^{\alpha}p_{\alpha}=m_e^2 c^2$:
\begin{align} \label{eq:mass-shell}
    \theta^\alpha_{~\alpha} = h c^2 \left[ m_e^2 - \left(\frac{n_0}{h}\right)^2 \right]                                   , \;
    N_\beta \theta^{\alpha\beta} = 0                 \; .
\end{align}
%
\section{Results}\label{sec:results}
%
In this section, we present a comparison between predictions of LEC and WARMC and also discuss their consistency w.r.t. PIC, by comparing the size of the first electron depletion bubble (\cref{subsec:bubble_size}) and the electromagnetic fields developed inside (\cref{subsec:electromagnetic_fields}). In all cases, we consider a background plasma with initial uniform density $n_i = 10^{16}~cm^{-3}$ (represented via $32$ particles-per-cell in the PIC simulations) and different values of the initial background temperature $T_i$. The driving electron bunch is modeled by a rigid, axially symmetric Gaussian density distribution, function of the radial coordinate $r$ and co-moving coordinate $\xi=z+ct$:
\begin{align}\label{eq:bunch-density}
    n_b(\xi,r) = \alpha \, \exp \left(-\frac{(\xi-\xi_0)^2}{2\sigma_z^2} - \frac{r^2}{2 \sigma_r^2} \right) \; ,
\end{align}
with rms widths $k_p \sigma_z = \sqrt{2}$, $k_p \sigma_r = 0.2$, and longitudinal center in $k_p \xi_0 = 6.0$, where $k_p = \omega_p / c$ is the cold plasma wave-number and $\omega_p = \sqrt{n_i e^2 / (m_e \epsilon_0)}$ is the cold plasma frequency. The bunch amplitude $\alpha$ was chosen to yield the desired value of the normalized charge parameter, $\tilde{Q}$, which controls the degree of non-linearity of the system:
\begin{align}
     \tilde{Q} = \left( \frac{\alpha}{n_i} \right) (k_p \sigma_z) (k_p \sigma_r)^2 (2\pi)^{3/2} \; .
\end{align}
While the fluid code is intrinsically axially symmetric, FBPIC was executed using only the fundamental azimuthal mode to enforce full cylindrical symmetry.
The computational domain is therefore represented by a two-dimensional grid $L_z \times L_r$, with $k_p L_z = 15$ and $k_p L_r = 5$, discretized with uniform resolution $k_p \Delta z = k_p \Delta r = 5\cdot10^{-3}$. The time step was set to $\omega_p \Delta t = 5\cdot10^{-4}$.
For convenience, the initial background temperature is sometimes expressed in this work in dimensionless form:
\begin{align}\label{eq:mui}
    \mu_i = \frac{k_B T_i}{m_e c^2} \;.
\end{align}

\subsection{Bubble Size}\label{subsec:bubble_size}
%
We first focus on the \textit{longitudinal} and \textit{transverse} bubble sizes, $\ell_{\parallel}$ and $\ell_{\bot}$, that provide a compact characterization of the geometry of the first electron depletion bubble formed in the plasma behind the driver~\cite{benedetti-2013,li-2015,stupakov-2016,wang-2017}. The two observables are here defined as follows (see~\cref{fig:1} for a sketch):
\begin{equation}
\begin{aligned}
    \ell_\parallel &= \xi_\parallel - \xi_m                  \;, \\ 
    \ell_\bot      &= r_m = r_b\left(\xi=\xi_m \right)    \;,
\end{aligned}
\end{equation}
where $\xi_\parallel$ is the point on the longitudinal axis where the bubble ends, $r_m$ is the maximum radial elongation of the bubble trajectory $r_b$, and $\xi_m$ is the corresponding co-moving coordinate. Before entering a detailed characterization of the bubble size in non-linear regimes, we took care in validating the numerical solvers used for the numerical simulations of LEC and WARMC. At difference with earlier validations already discussed in~\cite{simeoni-2024}, the validation proposed here specifically focuses on assessing the solver's capability to reproduce the geometry of the plasma bubble with high fidelity. We consider the linear regime, where $\tilde{Q} \ll 1$, allowing for analytical predictions for the plasma density in both LEC and WARMC. The theory has already been presented and used in~\cite{simeoni-2024} and is briefly reviewed in~\cref{sec:linear-theory}. In~\cref{fig:2} we analyze $\ell_{\parallel}$ and $\ell_{\bot}$ as a function of $k_B T_i$ for $\tilde{Q}=10^{-4}$. We observe that the numerical simulations for LEC and WARMC are in excellent agreement with the corresponding theoretical predictions, confirming the robustness of the numerical implementation used. These results for the bubble size in the linear regime will also be useful for comparisons with those obtained in the non-linear regimes (see below).
\begin{figure}[htbp!]
\includegraphics[width=\columnwidth]{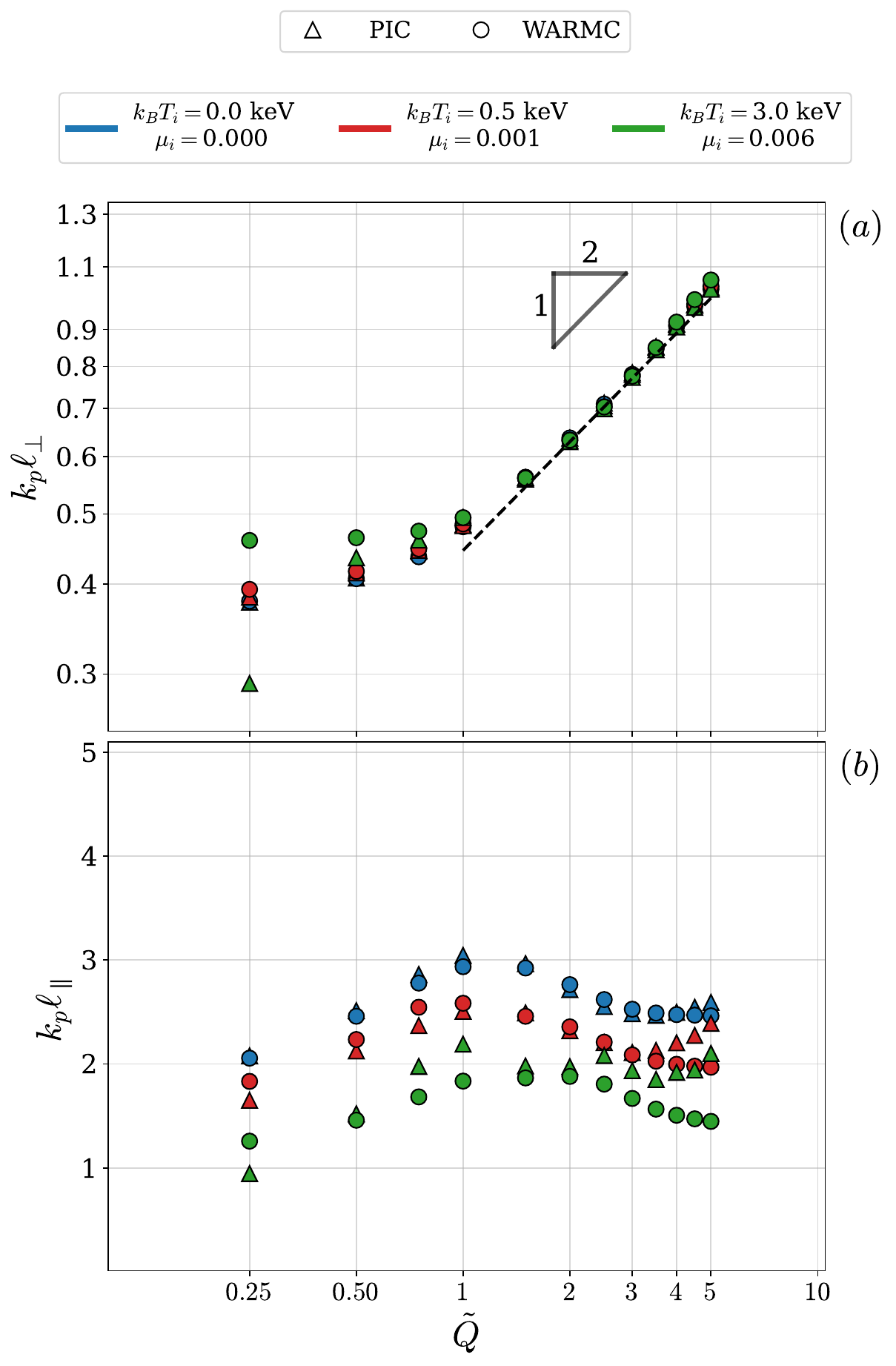}
    \caption{\footnotesize{We analyze $\ell_{\parallel}$ and $\ell_{\bot}$ at changing $\tilde{Q}$ for WARMC (circles) ad PIC (triangles) data for different $k_BT_i$: $k_BT_i=0~\rm{keV}$ (blue), $k_BT_i=0.5~\rm{keV}$ (red),
    $k_BT_i=3.0~\rm{keV}$ (green). Corresponding values of $\mu_i = k_B T_i/m_e c^2$ are indicated. Both $\ell_{\parallel}$ and $\ell_{\bot}$ are made dimensionless w.r.t. $k_p^{-1}$. Panel (a): log-log plot of $k_p \ell_\bot$ as a function of $\tilde{Q}$. We also report the scaling $k_p \ell_\bot \sim \tilde{Q}^{1/2}$ found in the cold limit (dashed black line, see text for details). Panel (b): lin-log plot of $k_p \ell_\parallel$ as a function of $\tilde{Q}$.} 
    }
    \label{fig:4}
\end{figure}
\begin{figure*}
\includegraphics[width=0.9\textwidth]{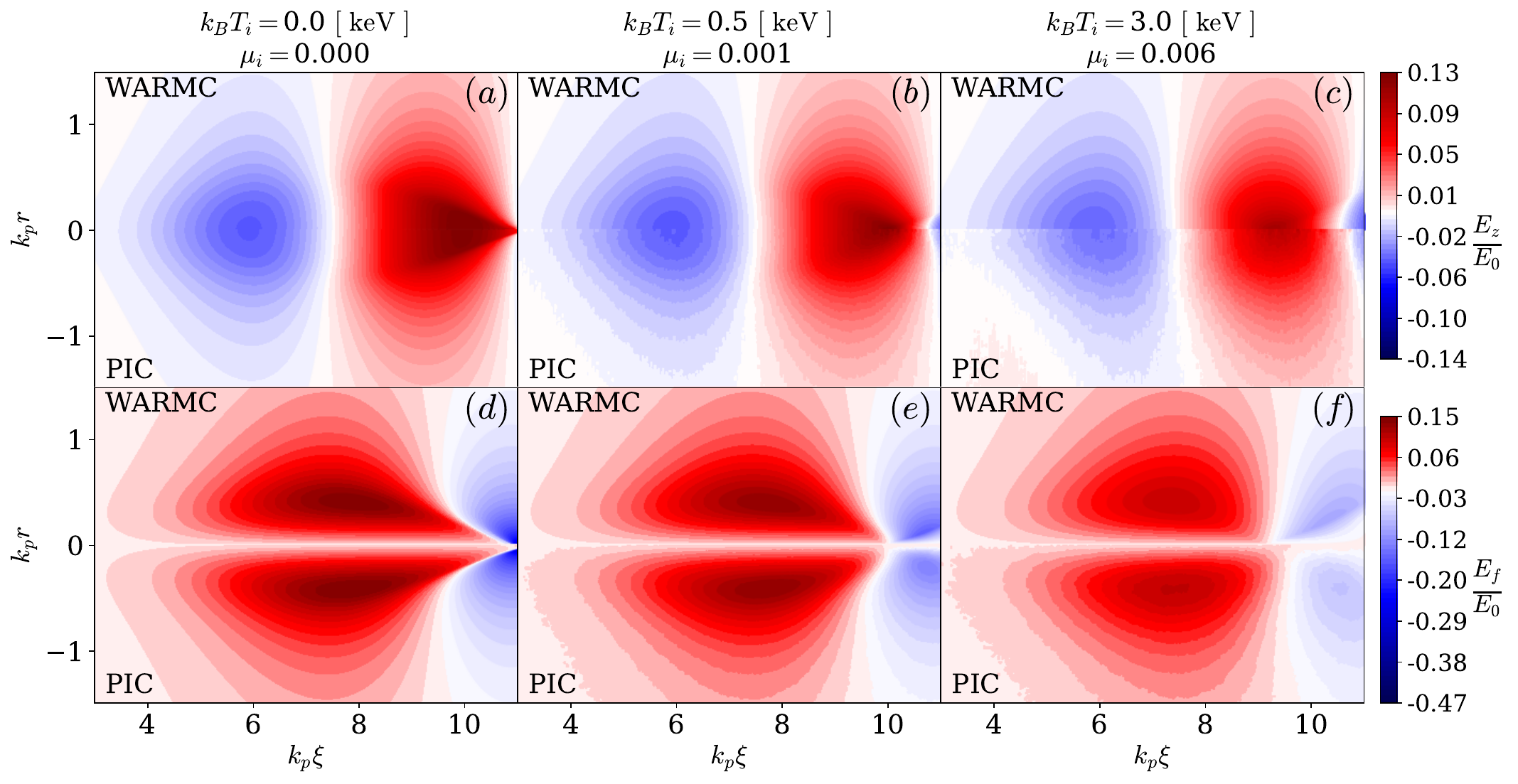}
    \caption{\footnotesize{Comparison between WARMC (top-half panels) and PIC (bottom-half) results for the accelerating field $E_z$ (top row, panels (a)-(c)) and the focusing field $E_f=E_r+cB_\phi$ (bottom row, panels (d)-(f)) for different $k_B T_i$: $k_BT_i = 0~\rm{keV}$ (first column, panels (a) and (d)), $k_BT_i = 0.5~\rm{keV}$ (second column, panels (b) and (e)), $k_BT_i = 3.0~\rm{keV}$ (third column, panels (c) and (f)). Corresponding values of $\mu_i = k_B T_i/m_e c^2$  are also indicated. Simulations are performed with $\tilde{Q}=1.0$. Fields and spatial coordinates are made dimensionless w.r.t. $E_0 = m_e c \omega_p / e$ and $k_p^{-1}$ respectively.}}
    \label{fig:5}
\end{figure*}
We then turn to the analysis at larger $\tilde{Q}$ by comparing the results of numerical simulations of the fluid models based on LEC and WARMC with those of PIC. Already from the snapshots reported in~\cref{fig:1}, it is possible to appreciate that, while in the cold limit both LEC and WARMC deliver results comparable with each
other and with the PIC data, in the presence of a finite temperature
($k_BT_i = 0.5$ keV, bottom half of all panels) the LEC yields a bubble geometry that diverges notably from the PIC prediction, whereas the WARMC prediction remains in closer agreement. A more quantitative insight is provided in~\cref{fig:3}, where in panels (a) and (b) we analyze $\ell_\parallel$ and $\ell_\bot$ as a function of $T_i$ for $\tilde{Q}=1.5$, corresponding to a sufficiently non-linear regime that enables the formation of a fully developed electron bubble. A direct comparison between the fluid and the PIC data reveals that the WARMC remains in closer agreement with the PIC data, providing a more accurate description of the bubble geometry than the LEC. This observation is consistent with the qualitative trends already highlighted in~\cref{fig:1}. Notably, the WARMC curves begin to deviate from the PIC reference around $k_BT_i \sim 1.0~\rm{keV}$ (corresponding to $T_i \sim 1.2 \cdot 10^7~\rm{K}$), which can therefore be interpreted as the upper validity limit of this closure. To our knowledge, this represents the first quantitative estimate of the regime of applicability of WARMC based on fully spatially resolved simulations compared with PIC simulations. Additionally, although the LEC exhibits larger discrepancies, panels (a) and (b) of~\cref{fig:3} also provide a measure of the extent of these deviations, offering useful guidance for users who may favor LEC reduced complexity over higher accuracy. We remark that both WARMC and LEC are formally derived under the assumption of low thermal spreads~\cite{simeoni-2024}, and hence deviations w.r.t. the reference PIC data are to be intended as early signatures of fluid models breakdown. The intensity of the discrepancies depends on the two geometric observables:
$\ell_\parallel$ is more sensitive to temperature because of the large electron density peak appearing on the longitudinal axis in the rear of the bubble, hence thermal smoothing appears more pronounced there. Conversely, thermal smoothing in the radial direction is less pronounced, and thus $\ell_\bot$ is less impacted. Notice that the variations of $\ell_{\bot}$ with temperature are more pronounced  in linear regimes (see~\cref{fig:2}): this indicates that non-linearity plays a significant role in shaping the thermal response of the bubble geometry~\cite{lu-2006,lu-2006-b}.\\
Having established that the WARMC provides the most accurate fluid description w.r.t. the PIC data, from now on, we will consider only this fluid closure in further analyses. To make progress, we then quantify how the accuracy of the WARMC depends on both $T_i$ and $\tilde{Q}$. To this aim, we analyze $\ell_{\parallel}$ as a function of $T_i$ for $\tilde{Q}$ in the range $1.0 \le  \tilde{Q} \le 4.5$ for both WARMC and PIC.  In fact, we chose to work with the longitudinal size because it is more sensitive to temperature w.r.t the transverse one in the range of $\tilde{Q}$ used, and hence it facilitates the analysis. For each $\tilde{Q}$, we determine the temperature $T_{\rm{max}}$ at which the relative discrepancy between the two models reaches a prescribed tolerance of $5\%$. The quantity $T_{\rm{max}}$ thus represents a threshold temperature beyond which the WARMC description ceases to provide quantitatively accurate predictions for the bubble geometry. The resulting values of $k_B T_{\rm{max}}$ against $\tilde{Q}$ are shown in panel (c) of~\cref{fig:3}: a monotonic decrease of the threshold temperature is observed at increasing $\tilde{Q}$. This signals that the validity range of the WARMC shrinks as the wake becomes more non-linear: in strongly non-linear regimes, even modest thermal spreads are sufficient to produce discrepancies w.r.t. the fully kinetic dynamics captured by PIC. This provides a compact and physically transparent way to assess the robustness of the fluid models across different operating conditions.
To complement the analysis of~\cref{fig:3}, we then investigated how $\ell_\parallel$ and $\ell_\bot$ depend on $\tilde{Q}$, and whether these dependencies are modified by a finite temperature. One could wonder, in fact, if the non-linear scalings of the blowout regime, established in the cold plasma limit~\cite{lu-2006,lu-2006-b,lindstrom-2025,golovanov-2017,mannan-2025}, are still valid when thermal effects are introduced. In~\cref{fig:4} we analyze $\ell_\bot$ and $\ell_\parallel$ as functions of $\tilde{Q}$ for three different initial background temperatures corresponding to $k_BT_i=0.0$, $0.5$, $3.0~\rm{keV}$.
Panel (a) shows that, from $\tilde{Q} \sim 1$ onward, data follow the $\ell_\bot \sim \tilde{Q}^{1/2}$ scaling that is peculiar of the blowout regime in the cold limit~\cite{lu-2006,lu-2006-b,mannan-2025,stupakov-2016,wang-2017,lindstrom-2025}. The scaling is not particularly affected by temperature, and remains clearly visible even at the highest $k_BT_i$ considered. More quantitatively, it is worth noting that the dashed line shown has exactly the form:
\begin{align}\label{eq:cold_theory_l_perp}
    k_p \ell_\bot = \frac{2.1 ~\tilde{Q}^{1/2}} {(2\pi)^{3/4} (k_p \sigma_z)^{1/2}} \;,
\end{align}
\begin{figure*}[h!]
    \centering
    \includegraphics[width=0.75\textwidth]{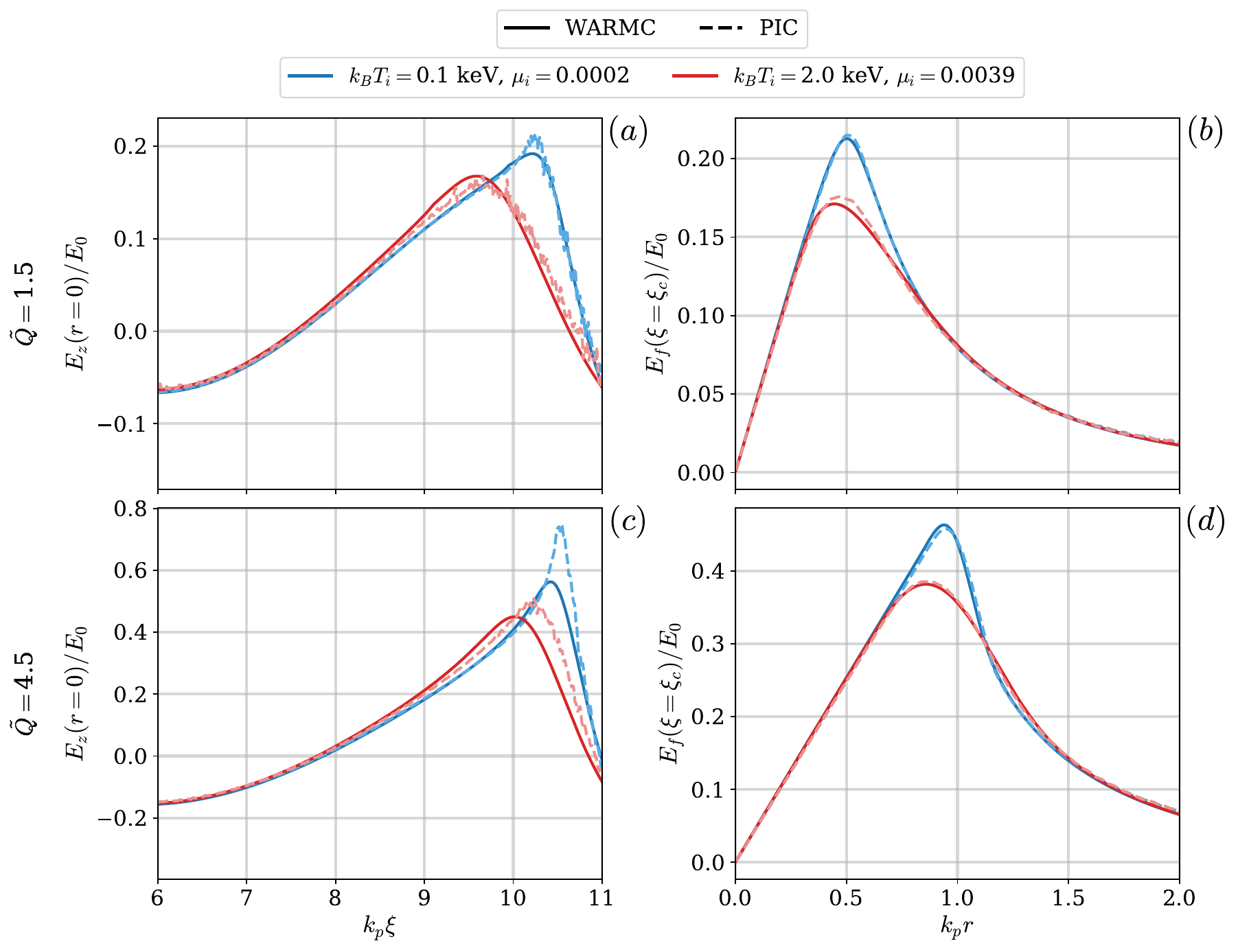}
    \caption{\footnotesize{Comparison between WARMC and PIC data for the accelerating field $E_z$ and the focusing field $E_f=E_r+cB_\phi$ for $\tilde{Q}=1.5$ (top row) and $\tilde{Q}=4.5$ (bottom row) and for different $k_B T_i$, $k_BT_i=0.1~\rm{keV}$ (blue) and $k_BT_i=2.0~\rm{keV}$ (red). Panels (a) and (c) show the on-axis accelerating field $E_z$ as a function of the longitudinal coordinate. Panels (b) and (d) display the focusing field $E_f$ as a function of the transverse coordinate. Fields and spatial coordinates are made dimensionless w.r.t. $E_0 = m_e c \omega_p / e$ and $k_p^{-1}$ respectively.}}
    \label{fig:6}
\end{figure*}
which aligns perfectly with the prediction found in~\cite{lu-2006,lu-2006-b} for elongated drivers ($k_p\sigma_z \sim 1$ and $k_p\sigma_r \ll 1$, as in our case), which only replaces the $2.1$ prefactor with a $2$. We further observe from panel (a) that for the largest values of $\tilde{Q}$ analyzed, $\ell_\bot$ lies in a regime where the bubble shape is expected to be close to an ellipse~\cite{lu-2006,lu-2006-b}. In this regime and in the cold limit, one expects that $\ell_\parallel$ can be reasonably approximated by half the plasma wavelength~\cite{mannan-2025}, i.e. $k_p \ell_{\parallel} \sim \pi$, in agreement with earlier results~\cite{lu-2006,stupakov-2016,wang-2017}. Indeed, panel (b) shows that $k_p \ell_\parallel$ in the cold limit shows a saturation to a value close to $\pi$ when $\tilde{Q}$ is close to 1. We observe a small overshoot that is found to depend on the adopted definition of $\ell_\parallel$: we have indeed verified that adjusting the definition of $\ell_\parallel$ as the distance between the centroid of the driver and the bubble rear yields a constant value devoid of the overshoot. Notice that for larger values of $\tilde{Q}$, we expect that $\ell_\parallel$ starts increasing and scaling proportionally to $\ell_{\bot}$~\cite{lu-2006,wang-2017}. These regimes are though not reachable by our fluid models and remain therefore unexplored. Regarding the impact of thermal effects, these appear to be more pronounced for $\ell_{\parallel}$ than for $\ell_{\bot}$: the values of $\ell_{\parallel}$ are actually lowered by temperature, although the functional dependency on $\tilde{Q}$ is not changed dramatically. This indicates that the underlying non-linear mechanism responsible for longitudinal saturation at $\tilde{Q}$ close to 1 is preserved and only shifted in magnitude. Finally, the deviations of WARMC predictions from PIC data are more pronounced at larger temperatures, consistently with the trends observed in~\cref{fig:3}.
%
\subsection{Electromagnetic fields}\label{subsec:electromagnetic_fields}
We now analyze the electromagnetic fields developed inside the bubble and we focus on both the accelerating field $E_z$ and the focusing field $E_f = E_r+cB_\varphi$. In~\cref{fig:5} we compare WARMC and PIC data, for three different values of the initial background temperature corresponding to $k_BT_i = 0.0, 0.5, 3.0~\rm{keV}$. The comparison shows that the validity limit previously identified for the WARMC also affects the electromagnetic fields, though its impact is milder, with noticeable discrepancies appearing mainly in the rear of the bubble. To render the analysis more quantitative, in~\cref{fig:6} we report longitudinal and transverse cuts of $E_z$ and $E_f$ respectively. The accelerating field is evaluated on-axis at $r=0$, while the focusing field is evaluated at the zero-crossing location $\xi=\xi_c$, defined as the point where $E_z$ changes sign. To quantitatively assess how the comparison between WARMC and PIC data changes at changing values of initial background temperature and degree of non-linearity, we report results for two values of $\tilde{Q}$ ($\tilde{Q}=1.5, 4.5$) and two values of $k_B T_i$ ($k_BT_i = 0.1, 2.0~\rm{keV}$). Panels (a) and (c) show that temperature acts on the slope $\partial_\xi E_z$ at the zero-crossing location via two competing mechanisms: first, a finite temperature smooths the field profile, reducing the peak of $E_z$ in the rear of the bubble and thus lowering the local derivative; second, thermal contraction shifts the field peak upstream, which increases $\partial_\xi E_z$. The figure shows that the second effect dominates, leading to an overall increase of the slope with temperature. By contrast, panels (b) and (d) show that the on-axis radial derivative of the focusing field $\partial_r E_f$ at the zero-crossing location is not affected by thermal effects. This is consistent with the milder temperature dependence expected for transverse quantities. Overall,~\cref{fig:6} reinforces the trends previously observed: when $T_i$ and $\tilde{Q}$ increase, the agreement between WARMC and the PIC data gradually deteriorates, with the largest deviations occurring in the rear of the bubble. Nonetheless, the disagreement remains moderate, and the WARMC closure continues to capture the essential field structure even at large temperatures. To further support these observations and to better assess how the field behavior varies across different acceleration regimes, we finally analyze $\partial_{\xi} E_z$ and $\partial_r E_f$, both evaluated at the zero-crossing location, in terms of $\ell_{\parallel}$ and $\ell_{\bot}$. Indeed, having previously established in the analysis of~\cref{fig:4} that the geometry of the bubble is close to an ellipse for the largest $\tilde{Q}$ analyzed, we can refer to scaling laws found for three dimensional ellipsoidal bunched beams~\cite{lapostolle-1965,wangler-2008}, predicting that the longitudinal derivative of the accelerating field scales linearly with the form-factor $f$:
\begin{align}\label{eq:form_factor}
    f = \frac{1}{3}\frac{\ell_\bot}{\ell_\parallel} \; .
\end{align}
\begin{figure}[h!]
\includegraphics[width=\columnwidth]{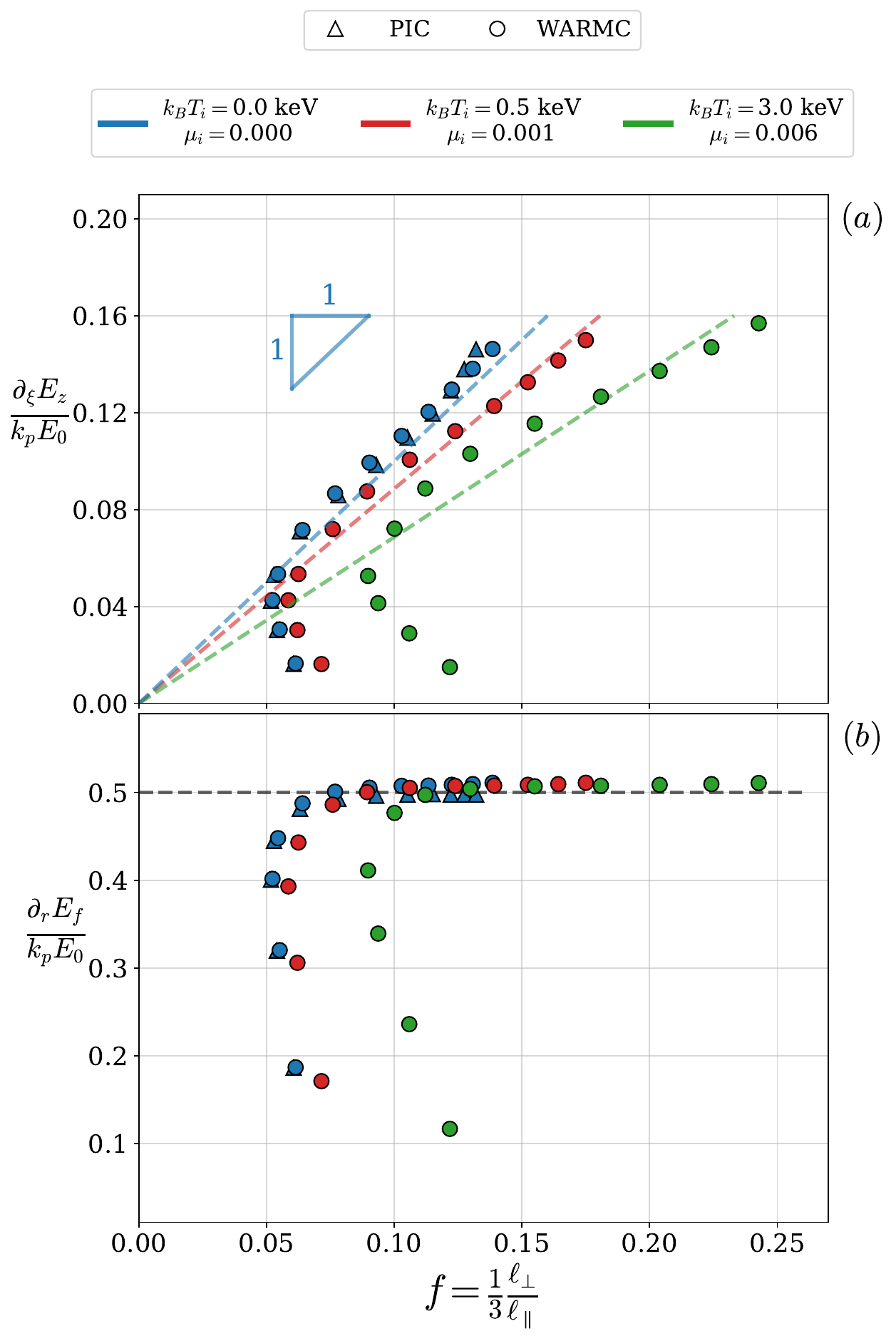}
    \caption{\footnotesize{Longitudinal and transverse fields derivatives evaluated at $(r=0,\xi=\xi_c)$ as functions of the form-factor $f$ (see text for details), for different $k_B T_i$: $k_B T_i = 0.0~\rm{keV}$ (blue), $k_B T_i = 0.5~\rm{keV}$ (red), $k_B T_i = 3.0~\rm{keV}$ (green). We report both PIC (triangles) and WARMC (circles) data. Panel (a) shows the longitudinal derivative of the accelerating field. We also report the linear scaling predicted by the cold theory (dashed blue line, see text for details) and fitted linear scaling laws for the two warm cases (dashed red and green lines). Panel (b) shows the radial derivative of the focusing field. The black dashed line denotes the value $1/2$. Derivatives are made dimensionless w.r.t. $k_p E_0$, where $E_0 = m_e c \omega_p / e$.}
    }
    \label{fig:7}
\end{figure}
Hence, in~\cref{fig:7} we analyze $\partial_{\xi} E_z$ and $\partial_r E_f$ as a function of $f$ for three different initial background temperatures corresponding to $k_BT_i=0.0$, $0.5$, $3.0~\rm{keV}$. Panel (a) of~\cref{fig:7} shows excellent agreement between the cold PIC data, the cold fluid data, and the predicted linear scaling (dashed blue line in the panel), except for those points that sit at the lowest derivative values and correspond to small $\tilde{Q}$, and therefore lie outside the ellipsoidal regime~\cite{lu-2006,lu-2006-b}. In the presence of a non-negligible temperature, we preferred not to consider PIC data, as numerical differentiation in the presence of PIC noise would require post-processing smoothing filters that could introduce artificial modifications and compromise the physical interpretation of the results. Regarding WARMC data, when $T_i>0$, we still observe linear scalings with the form-factor $f$ (dashed red and green lines in the panel), although with smaller slopes ($0.8$ for $k_B T_i = 0.5~\rm{keV}$ and $0.6$ for $k_B T_i = 3.0~\rm{keV}$). This probably signals that a geometric interpretation~\cite{lapostolle-1965,wangler-2008} for the electromagnetic fields in the bubble is still in place even in the warm cases, although the reduced slopes might represent the fact that in these cases the bubble has less sharp boundaries. Lastly, panel (b) highlights that the transverse derivative approaches a plateau at large $f$, with an asymptotic value of $1/2$, consistent with previous observations in similar regimes~\cite{rosenzweig-1991} and theoretical models~\cite{lu-2006,lu-2006-b}. We observe that the asymptotic value is reached for growing values of $f$ as the temperature increases. Nevertheless, in both cases, thermal effects produce only minor modifications, confirming that these fields are roughly independent of the considered temperature range.

\section{Conclusions}\label{sec:conclusions} 
%
We have characterized the wakefield structure in the process of plasma acceleration with a non-negligible initial background temperature $T_i$ and for different degrees of non-linearity in the wake, parametrized via the normalized charge parameters $\tilde{Q}$. We quantitatively assessed the validity of fluid models with two different popular thermal closure assumptions, the LEC~\cite{toepfer-1971} and WARMC~\cite{schroeder-2005,schroeder-2009,schroeder-2010}, when $T_i$ and $\tilde{Q}$ are varied. Results from fluid models were compared against PIC simulations performed with the code FBPIC~\cite{lehe-2016}. To this end, we have analyzed a set of physically relevant observables: the longitudinal ($\ell_\parallel$) and transverse ($\ell_{\bot}$) size of the first electron depletion bubble formed in typical wakefield acceleration setups~\cite{benedetti-2013,li-2015,stupakov-2016,wang-2017}, as well as the accelerating and focusing fields, which directly determine the acceleration performance and for which 
well-established scaling laws exist~\cite{lu-2006,lu-2006-b,lapostolle-1965,wangler-2008}.\\
Our analysis delivers a series of key findings. First, among the two thermal closure assumptions, WARMC provides the most accurate description of the plasma wake as $T_i$ varies, with significantly improved predictions for both $\ell_\parallel$ and $\ell_\bot$, compared with the LEC. WARMC also reproduces the accelerating and focusing fields with good fidelity, except for the region in the rear of the bubble at the largest temperatures considered. Second, by analyzing the temperature at which the discrepancy between PIC and WARMC reaches a prescribed threshold, we have identified the temperature range where WARMC remains reliable. This validity range is found to get narrower as $\tilde{Q}$ increases, reflecting that the validity of fluid models -- that are in fact derived on the assumption of small thermal spread -- decreases in strongly non-linear regimes. These limits have to be taken into consideration when applying fluid modelling in contexts where thermal effects ranging from sub-$\rm{keV}$ to a few $\rm{keV}$ can influence the plasma response~\cite{gholizadeh-2011,silva-2021,diederichs-2023,diederichs-2023-b,cao-2024,zgadzaj-2020,khudiakov-2022}.
Third, our study highlights that thermal effects become more evident in the longitudinal observables, in both geometric and field based observables. The transverse observables are indeed less sensible to thermal effects, and maintain qualitatively the same scaling laws peculiar of the cold limit~\cite{lu-2006,lu-2006-b,lapostolle-1965,wangler-2008}. \\
Our findings suggest several directions for future work. A first natural extension is to investigate whether the WARMC closure can be systematically improved by including higher-order moments of the Vlasov hierarchy. Such an approach may extend its range of validity to larger thermal spreads, although at the cost of increased model complexity. 
Second, this work motivates the exploration of hybrid fluid–PIC strategies, in which a fluid description is employed where valid, while PIC is activated only in localized regions or time intervals where the fluid model demonstrably breaks down. Such adaptive schemes may provide an efficient and accurate alternative to fully kinetic simulations in high-energy plasma accelerator modeling with reduced numerical noise.


\section*{Acknowledgments}
The authors gratefully acknowledge Fabio Bonaccorso for his technical support. This work was supported by the Italian Ministry of University and Research (MUR) under the FARE program (No. R2045J8XAW), project "Smart-HEART". MS gratefully acknowledges the support of the National Center for HPC, Big Data and Quantum Computing, Project CN\_00000013 - CUP E83C22003230001, Mission 4 Component 2 Investment 1.4, funded by the European Union - NextGenerationEU.


\section*{Author Declarations}

\subsection*{Conflict of Interests}
The authors declare that they have no conflict of interests.

\subsection*{Author Contributions}
\textbf{Daniele Simeoni}: 
Conceptualization (equal); 
Data Curation (lead); 
Formal Analysis (lead); 
Investigation (lead); 
Software (lead); 
Visualization (lead); 
Writing - original draft (equal). 
\textbf{Andrea Renato Rossi}: 
Conceptualization (equal); 
Investigation (supporting); 
Writing - original draft (supporting).
\textbf{Gianmarco Parise}: 
Conceptualization (supporting); 
Software (supporting); 
Writing - original draft (supporting).
\textbf{Fabio Guglietta}: 
Conceptualization (supporting); 
Software (supporting); 
Writing - original draft (supporting).  
\textbf{Mauro Sbragaglia}: 
Conceptualization (equal); 
Writing - original draft (equal).


\section*{Data Availability Statement}

The data that support the findings of this study are available from the corresponding author upon reasonable request.


\appendix

\section{Theory for linearized fluid models}\label{sec:linear-theory}
In the presence of weak driving bunches, it can be assumed that the plasma is linearly perturbed from the initial rest state. In these cases, it is possible to apply perturbation theory to the fluid equations in both closures, respectively~\cref{eq:rel-euler} and~\cref{eq:warmc-fluid-eqs}, coupled with Maxwell~\cref{eq:maxwell-homogeneus,eq:maxwell-inhomogeneus}, obtaining in this way an equation for the density perturbation $\tilde{n}=n-n_i$:
\begin{align}\label{eq:fhow}
    \left[ \partial_t^2  - c_s^2 \nabla^2 + a_e^2 \right] \tilde{n} = - a_b^2 n_b \;,
\end{align}
that is a forced Klein-Gordon equation with coefficients depending on the selected closure scheme and the initial background temperature of the plasma. At the first order in $T_i$ one gets:
\[
\renewcommand{\arraystretch}{1.3}
\begin{array}{l|c|c}
                    & \textbf{LEC} & \textbf{WARMC}             \\ 
  \hline
  (c_s/c)^2         & \tfrac{5}{3}\mu_i & 3\mu_i                \\
  \hline
  (a_e/\omega_p)^2  & 1-\tfrac{5}{2}\mu_i & 1-\tfrac{5}{2}\mu_i \\
  \hline
  (a_b/\omega_p)^2  & 1-\tfrac{5}{2}\mu_i & 1-\tfrac{1}{2}\mu_i
\end{array}
\]
One then assumes that the longitudinal dynamics of the system is co-moving with the driving bunch propagating along the z-direction. Thus, every field depends on the co-moving coordinate $\xi$.~\cref{eq:fhow} then becomes:
\begin{align}\label{eq:fhow-comoving}
    \left[ 
      (c^2 -c_s^2) \partial_\xi^2  
    - c_s^2 \nabla_{\bot}^2 
    + a_e^2 \right] \tilde{n} 
    = - a_b^2 n_b \; ,
\end{align}
where $\nabla_{\bot}^2$ denotes the transverse Laplacian operator. Furthermore, due to the cylindrical geometry of the problem, one assumes the density $n_b$ to be axially symmetric, and therefore to be dependent only on the transverse radial coordinate $r$ and the co-moving variable $\xi$.
It then becomes convenient to apply to~\cref{eq:fhow-comoving} the Hankel transform of order 0~\cite{offord-1935}:
\begin{align}
     H_0 \left[ f(r) \right] = \hat{f}(w) = \int_{0}^{+\infty} r f(r) J_0 (w r) dr  \; ,
\end{align}
where $J_0 (wr)$ is the Bessel function of first kind and order 0. Under this transform, the transverse Laplacian becomes $ \nabla_{\bot}^2 \rightarrow -w^2$, so one obtains a forced harmonic oscillator in the variable $\xi$:
\begin{align}
     \left[ \partial_\xi^2 + k_w^2 \right] \hat{n}(w,\xi) &= f_b \hat{n}_b(w,\xi) \; , \\
     k_w^2 = \frac{a_e^2 + w^2 c_s^2}{c^2-c_s^2}                                     &, \;
       f_b = -\frac{a_b^2}{c^2-c_s^2}                                                  \; ,
\end{align}
which can be formally solved via the Green's function method:
\begin{align}
      \hat{n}(w,\xi) &= f_b \int_{-\infty}^{\xi} \hat{n}_b(w,\xi') \frac{\sin[k_w(\xi-\xi')]}{k_w} d\xi' \\
    \tilde{n}(r,\xi) &= \int_{0}^{\infty} \hat{n}(w,\xi) J_0(wr) w dw  .    
\end{align}\label{eq:fhow-solution}
The above expressions can be numerically evaluated to obtain the linear density perturbation $\tilde{n}$ for both LEC and WARMC.

\printbibliography

\end{document}